\documentclass[conference]{IEEEtran}
\IEEEoverridecommandlockouts

\usepackage{cite}
\usepackage{amsmath,amssymb,amsfonts}
\usepackage{algorithm}
\usepackage{algpseudocode}
\usepackage{amsmath}
\usepackage{graphicx}
\usepackage{textcomp}
\usepackage{footnote}
\usepackage{hyperref}
\usepackage[absolute,overlay]{textpos}
\usepackage[usenames]{color}
\usepackage{array}
\usepackage{float}
\usepackage{xcolor}
\usepackage{pifont} 
\usepackage{multirow} 
\usepackage{subcaption} 
\usepackage{svg}
\usepackage{booktabs}
\usepackage[none]{hyphenat}
\def\BibTeX{{\rm B\kern-.05em{\sc i\kern-.025em b}\kern-.08em
    T\kern-.1667em\lower.7ex\hbox{E}\kern-.125emX}}
\begin{document}

\title{
Impedance-Based Sensitivity Analysis for Stability Enhancement of LCC-HVDC Links Connected to Weak Grids Using Grid-Forming Converters
% Impedance-Based Sensitivity Analysis for \\ Stability Enhancement of LCC-HVDC \\ Systems Using Grid-Forming Converters
% Sensitivity‑Driven Impedance Analysis for Stability Improvement of LCC‑HVDC Systems with Grid‑Forming Converters
% \\
% {\footnotesize \textsuperscript{*}Note: Sub-titles are not captured for https://ieeexplore.ieee.org  and
\thanks{
% This work has received funding from the ADOreD project under the European Union’s Horizon Europe Research and Innovation Programme under the Marie Skłodowska-Curie Grant Agreement No. 101073554. \\
$^{\dagger}$The corresponding author.}
}

\author{\IEEEauthorblockN{Luis A. Garcia-Reyes$^{\dagger}$}
\IEEEauthorblockA{\textit{CITCEA-UPC} \\
Barcelona, Spain \\
luis.reyes@upc.edu}
\and
\IEEEauthorblockN{Anup Joshi}
\IEEEauthorblockA{\textit{L2EP - Centrale Lille} \\
Lille, France \\
anup.joshi@centralelille.fr}
\and
\IEEEauthorblockN{Javier Renedo}
\IEEEauthorblockA{\textit{Red Eléctrica - Redeia} \\
Madrid, Spain \\
javier.renedo@ree.es}
\and
\IEEEauthorblockN{Macarena Martin-Almenta}
\IEEEauthorblockA{\textit{Red Eléctrica - Redeia} \\
Madrid, Spain \\
macarena.martin@ree.es}
\and
\IEEEauthorblockN{Oriol Gomis-Bellmunt}
\IEEEauthorblockA{\textit{CITCEA-UPC} \\
Barcelona, Spain \\
oriol.gomis@upc.edu}
\and
\IEEEauthorblockN{Eduardo Prieto-Araujo}
\IEEEauthorblockA{\textit{CITCEA-UPC} \\
Barcelona, Spain \\
eduardo.prieto-araujo@upc.edu}
\and
\IEEEauthorblockN{Vinicius A. Lacerda}
\IEEEauthorblockA{\textit{CITCEA-UPC} \\
Barcelona, Spain \\
vinicius.lacerda@upc.edu}
\and
\IEEEauthorblockN{Xavier Guillaud}
\IEEEauthorblockA{\textit{L2EP - Centrale Lille} \\
Lille, France \\
xavier.guillaud@centralelille.fr}
}
% \vspace{1pt}
% \begin{textblock*}{\textwidth}(1.8cm,25.9cm)
% \noindent
% {\fontsize{8}{9.5}\selectfont
% 979-8-3315-2503-3/25/\$31.00~\copyright~2025 IEEE}
% \end{textblock*}
% \vspace{-6mm}
\maketitle

% \vspace{-12mm}

\begin{abstract}
This paper presents a frequency-domain, impedance-based sensitivity methodology for stability assessment and enhancement of line-commutated converter HVDC (LCC-HVDC) links operating under weak-grid conditions. The methodology integrates frequency-domain identification tailored for black-box systems, the Generalized Nyquist Criterion (GNC) for multivariable stability assessment, and modal impedance decomposition with participation-factor analysis to locate and interpret interaction mechanisms. The approach is validated against a detailed linearized state-space model and nonlinear EMT simulations of an LCC-HVDC benchmark. A sensitivity study varying the grid short-circuit ratio (SCR) reveals a stability limit for the standalone LCC-HVDC link and demonstrates that the integration of a grid-forming voltage source converter (GFM-VSC) substantially increases the stability margin.
\end{abstract}

\vspace{-3mm}
\section{Introduction}

The increasing penetration of renewable generation, notably offshore and onshore wind, is driving wider deployment of high‑voltage direct‑current (HVDC) links to enhance capacity, flexibility and controllability in long‑distance transmission \cite{oriol,paper_Mario}. Line‑commutated converter HVDC (LCC‑HVDC) remains a mature and high‑power solution for bulk transfer and asynchronous alternating current (AC) interconnections, often enabling larger power transfers than voltage source converter HVDC (VSC‑HVDC) alternatives \cite{paper_overview_VSC-Hvdc,paper_comparative_LCC_HVDC,paper_comparative_HVDC_classic}. At the same time, rising renewable penetration reduces system strength and short‑circuit capability, increasing LCC‑HVDC susceptibility to commutation failures because LCCs cannot supply reactive power during disturbances \cite{paper_commutation_failure}. Static synchronous compensators (STATCOMs) and other VSC‑based supports have been proposed to mitigate these dynamics, but their integration can introduce sub‑ and super‑synchronous interactions in weak grids, creating new stability and interaction challenges \cite{paper_STATCOM_LCC-HVDC,paper_GFM_LCC}.

Small‑signal analysis (SSA) based on state‑space eigenvalue methods is the standard tool to identify instability mechanisms and interactions \cite{paper_CCM2}. However, detailed SSA requires access to converter internal models that are often unavailable due to manufacturer intellectual‑property (IP) restrictions, yielding black‑box devices and limiting applicability in multi‑vendor environments. To overcome this, SSA impedance‑based (SSA‑IB) using frequency‑domain (FD) identification has been used to characterize black‑box systems and perform SSA studies using the Generalized Nyquist Criterion (GNC) \cite{paper_molinas_methodology_mimo_siso,paper_nyquist}. Despite its appeal, SSA‑IB faces two main challenges for LCC‑HVDC: (i) the strong nonlinearities and frequency coupling introduced by thyristor commutation, which complicate linear identification \cite{paper_harmonic_ss_LCC,paper_agusti_LCC}, and (ii) the need for FD identification validation that operates reliably in electromagnetic transient (EMT) environments and does not depend on access to detailed SSA models \cite{paper_non-intrusive_LCC,paper_scanning_LCC,paper_SIaD}.

Several works have advanced SSA and impedance modeling for LCC‑HVDC. High‑fidelity state‑space SSA formulations capture commutation dynamics and enable sensitivity studies when detailed models are available \cite{paper_accurate_SSA-LCC,paper_comparative_SSA-LCC}. Impedance models derived from SSA improve representation of frequency coupling and commutation effects \cite{paper_IB_modeling_LCC,paper_agusti_LCC,paper_harmonic_ss_LCC}, while non‑intrusive wideband identification techniques and scanning tools provide practical means to obtain black‑box impedances \cite{paper_non-intrusive_LCC,paper_scanning_LCC,paper_SIaD}. Separately, grid-forming (GFM) VSCs have been proposed to enhance stability of LCC-HVDC links connected to weak grids, analyzing interactions between GFM‑VSCs and LCC links and demonstrating potential damping benefits \cite{paper_STATCOM_LCC-HVDC,paper_Anup}. However, these contributions are based on detailed models or do not fully address validation and applicability in multi‑vendor scenarios.

This paper bridges these gaps by introducing a frequency‑domain (FD), impedance‑based sensitivity methodology tailored to LCC‑HVDC systems operating under weak‑grid conditions. The approach (i) obtains accurate impedance representations of both the LCC‑HVDC link and the AC grid via FD identification using an open‑source tool \cite{paper_SIaD}, (ii) applies the GNC for multivariable stability assessment, and (iii) employs modal impedance decomposition and participation factors to identify and quantify potential sub‑ and super‑synchronous interactions. The impedance models are validated against the linearized state‑space and nonlinear EMT models introduced in \cite{paper_Anup}. A parametric study varying the short‑circuit ratio (SCR) quantifies stability limits for the standalone LCC‑HVDC link and demonstrates that integrating a grid‑forming VSC (GFM‑VSC), operating as an enhanced STATCOM or a battery energy storage interface, substantially increases the stability margin of the system.

\section{Impedance-Based Methodology Analysis}

\begin{figure}[t!]
    \centering
    % Image a)
    \begin{subfigure}[b]{0.75\columnwidth}
        \centering
        \includegraphics[width=\textwidth]{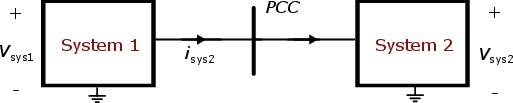}
        \vspace{-5mm}
        \caption{Point under study in the system.}
        \vspace{2mm}
        \label{fig:impedance-based1}
    \end{subfigure}
    % \vspace{1em} % Space between images
    % Image b)
    \begin{subfigure}[b]{0.78\columnwidth}
        \centering
        \includegraphics[width=\textwidth]{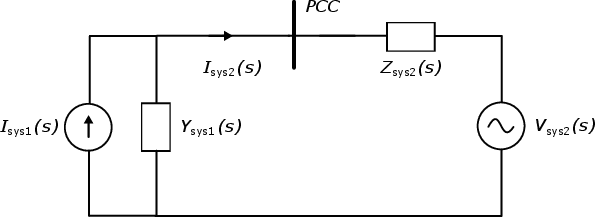}
        \vspace{-5mm}
        \caption{Equivalent source-load model.}
        \vspace{2mm}
        \label{fig:impedance-based2}
    \end{subfigure}
    % imagen c)
    \begin{subfigure}[b]{0.78\columnwidth}
        \centering
        \includegraphics[width=\textwidth]{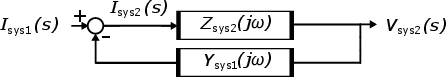}
        \vspace{-5mm}
        \caption{Block diagram of the transfer function.}
        \label{fig:impedance-based3}
    \end{subfigure}
    \vspace{-1mm}
    \caption{Impedance-based stability model.}
    \label{fig:impedance_based}
    \vspace{-5mm}
\end{figure}

As illustrated in Fig.~\ref{fig:impedance_based}, a power system can be equivalently represented in two subsystems for SSA-IB analysis in the FD \cite{paper_molinas_methodology_mimo_siso,paper_SIaD}. In Fig.~\ref{fig:impedance-based1}, the complete system is shown, and at the Point of Common Coupling (PCC), a source–load equivalent can be employed, as depicted in Fig.~\ref{fig:impedance-based2}, where one subsystem is represented by a Norton equivalent and the other by a Thévenin equivalent. Fig.~\ref{fig:impedance-based3} presents the block diagram of the closed-loop response, whose mathematical relationship is
\begin{equation}
I_{sys2}(s)= \frac{ I_{sys1}(s) - Y_{sys1}(s) V_{sys2}(s) }{1+Y_{sys1}(s) Z_{sys2}(s)},
\end{equation}
where $I_{sys1}(s)$ and $I_{sys2}(s)$ denote the currents of subsystems 1 and 2, respectively; $V_{sys2}(s)$ is the voltage of subsystem 2; and $Y_{sys1}(s)$ and $Z_{sys2}(s)$ are the admittance and impedance of the two subsystems. Here, $s=j\omega$ is the Laplace operator, with $\omega$ being the angular frequency in rad/s. Note that instability arises when the denominator becomes undefined.

In Multiple-Input Multiple-Output (MIMO) systems, the stability criterion used is the Generalized Nyquist Criterion (GNC), which evaluates the minor-loop gain defined as
\begin{equation}\label{eq:Lmatrix}
\mathbf{L}(s)=\mathbf{Y}_{\text{sys1}}(s)\,\mathbf{Z}_{\text{sys2}}(s),
\end{equation}
where the $dq$-frame matrices are expressed as
\begin{subequations}
\begin{equation}
\mathbf{Y}_{\text{sys1}}(s)=
\begin{bmatrix}
Y_{sys1}^{dd}(s) & Y_{sys1}^{dq}(s) \\
Y_{sys1}^{qd}(s) & Y_{sys1}^{qq}(s)
\end{bmatrix},
\end{equation}
\begin{equation}
\mathbf{Z}_{\text{sys2}}(s)=
\begin{bmatrix}
Z_{sys2}^{dd}(s) & Z_{sys2}^{dq}(s) \\
Z_{sys2}^{qd}(s) & Z_{sys2}^{qq}(s)
\end{bmatrix}.
\end{equation}
\end{subequations}

The GNC assesses the trajectories of the eigenvalues of $\mathbf{L}(s)$ and states that stability is ensured when none of them encircle the critical point $(-1,j0)$. The eigenvalues are obtained from
\begin{equation}
\det(\mathbf{I}+\mathbf{L}(s))=\prod_i (1+\lambda_i(s)),
\end{equation}
where $\lambda_i(s)$ denotes the $i$-th eigenvalue of \eqref{eq:Lmatrix}.

In modern power systems, due to model complexity and IP restrictions, $\mathbf{Y}_{\text{sys1}}(s)$ and $\mathbf{Z}_{\text{sys2}}(s)$ are typically obtained through FD scanning techniques \cite{paper_SIaD}. Although identification can be performed in $abc$, $dq$, or $pn$ sequences, this work focuses on the $dq$ frame, which simplifies the analysis while accurately capturing small-signal phenomena \cite{paper_agusti_LCC,paper_SIaD}.

Once the subsystem matrices are identified, the oscillatory behavior of the complete system is determined as in \cite{paper_SIaD}
\begin{equation}
\mathbf{Y}_{\text{sys}}(s)=\mathbf{Y}_{\text{sys1}}(s)+\left[\mathbf{Z}_{\text{sys2}}(s)\right]^{-1}.
\end{equation}

The modal decomposition is given by
\begin{subequations}
\begin{equation}
\mathbf{Y}_{\text{sys}}(s)=\mathbf{\Upsilon}(s)\,\mathbf{\Lambda}(s)\,\mathbf{\Phi}(s),
\end{equation}
\begin{equation}
\mathbf{Z}_{mode}(s)=\left[\mathbf{\Lambda}(s)\right]^{-1},
\end{equation}
\end{subequations}
where $\mathbf{\Upsilon}(s)$ and $\mathbf{\Phi}(s)$ are the right and left eigenvector matrices, $\mathbf{\Lambda}(s)$ is the diagonal eigenvalue matrix, and $\mathbf{Z}_{mode}(s)$ denotes the modal impedances.

Finally, participation factors are computed as
\begin{equation}
P_{ki}(s)=\frac{|\Upsilon_{ki}(s)\,\Phi_{ik}(s)|}{\sum_{k=1}^{N}|\Upsilon_{ki}(s)\,\Phi_{ik}(s)|},
\end{equation}
allowing identification of the contribution of each input to the $i$-th mode \cite{paper_SIaD}.

\section{Impedance-Based Sensitivity Analysis}

\begin{figure}[t!]
% \vspace{-10mm}
    \centering
    \begin{subfigure}[b]{0.8\columnwidth}
        \centering
        \includegraphics[width=0.95\textwidth]{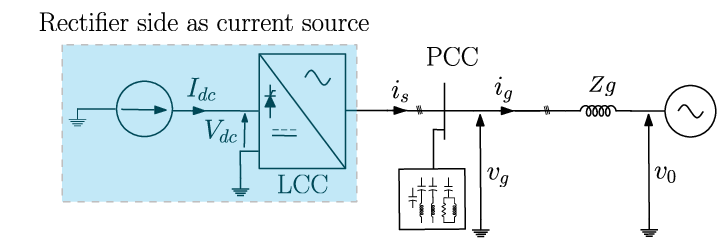}
        % \vspace{-10mm}
        \caption{LCC-HVDC link connection model detail.}
        \label{fig:detailed_LCC}
    \end{subfigure}
    \begin{subfigure}[b]{0.8\columnwidth}
    % % \vspace{-9mm}
        \centering
        \includegraphics[width=0.95\textwidth]{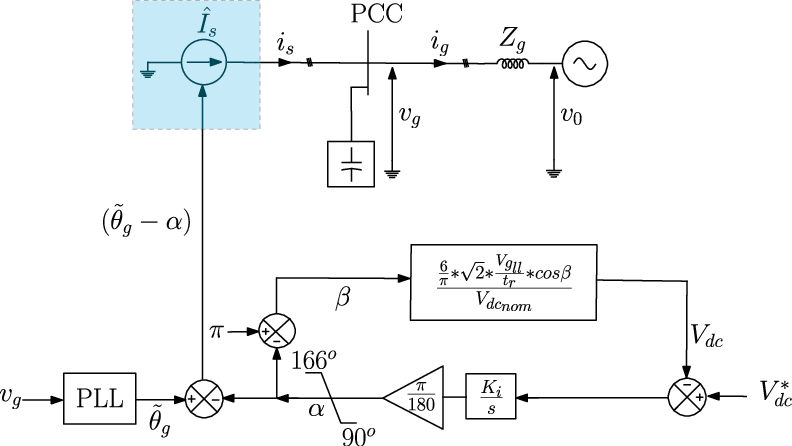}
        % \vspace{-10mm}
        \caption{Simplified model.}
        \label{fig:simplified_LCC}
    \end{subfigure}
    \vspace{-1mm}
    \caption{LCC-HVDC link connected to a grid equivalent.}
    \vspace{-1mm}
    \label{fig:LCC}
\end{figure}

For the analysis presented in this paper, the system described in \cite{paper_Anup} is considered. The rectifier is modeled as a constant DC current source and is connected to the AC side through an ideal three-winding $Y\!-\!\Delta$ transformer, as shown in Fig.~\ref{fig:detailed_LCC}. For the study, a simplified averaged LCC-HVDC link is adopted: a three-phase controlled current source interfaced to a grid equivalent, together with an ideal DC voltage control that represents the plant dynamics (see Fig.~\ref{fig:simplified_LCC}). This reduced model accurately reproduces the dynamics of the detailed switching LCC-HVDC model analyzed in \cite{paper_Anup}. More details on the model assumptions and control implementation are available in \cite{paper_Anup}. System parameters for the full LCC-HVDC model are listed in Tables \ref{table:LCC_parameters} and \ref{table:Vdc_parameters}.

\begin{table}[t!]
\caption{Parameters of the system under study.}
\label{table:LCC_parameters}
\centering
\begin{tabular}{ccc}
\toprule
\textbf{Component} & \textbf{Parameter} & \textbf{Value} \\
\midrule
\multirow{3}{*}{\parbox[c]{2.8cm}{\centering LCC-HVDC}} 
  & $V_{dc,\mathrm{nom}}$ & 500\;kV \\
  & $I_{dc,\mathrm{nom}}$ & 2000\;A \\
  & $P_{n,\mathrm{LCC}}$  & 1000\;MW \\
\addlinespace
Capacitor bank
  & $Q_c$                 & 600\;Mvar \\
\addlinespace
\multirow{2}{*}{\parbox[c]{2.8cm}{\centering AC grid}} 
  & $U_n$                 & 400\;kV \\
  & SCR                 & 10 \\
\bottomrule
\end{tabular}
\vspace*{-1mm}
\end{table}

\begin{table}[t!]
\caption{Control parameters of the LCC-HVDC.}
\label{table:Vdc_parameters}
\centering
\begin{tabular}{ccc}
\toprule
\textbf{Control loop} & \textbf{Parameter} & \textbf{Value} \\
\midrule
\multirow{2}{*}{\parbox[c]{3.2cm}{\centering DC voltage control}}
  & $T^{5\%}_r$ & 300\;ms \\
  & $k_i$                       & 1000\;deg/pu/s \\
\addlinespace
\multirow{2}{*}{\parbox[c]{3.2cm}{\centering Phase locked loop (PLL)}}
  & $\omega_n$ & 50\;rad/s \\
  & $\zeta$    & 0.707 \\
\bottomrule
\end{tabular}\vspace{-4mm}
\end{table}

\begin{figure}[t!]
\centering
\includegraphics[width=0.75\columnwidth]{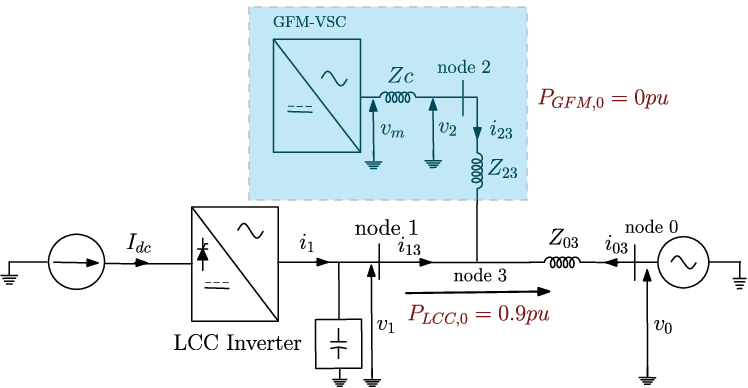}
\caption{Scheme of the connection of a GFM-VSC to the LCC-HVDC system for stability enhancement.}
\label{fig:GFM_HVDC}
\vspace{-4mm}
\end{figure}

\begin{figure}[t!]
% \vspace{-10mm}
    \centering
    \begin{subfigure}[b]{0.75\columnwidth}
        \centering
        \includegraphics[width=0.85\textwidth]{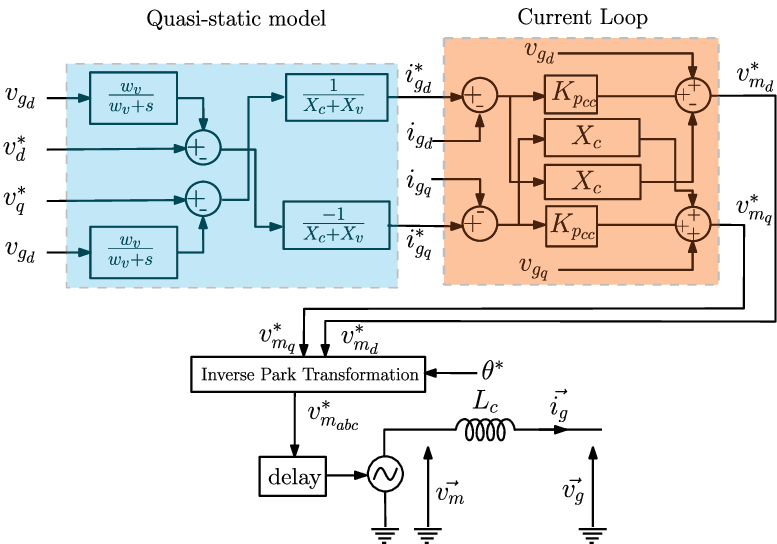}
        % \vspace{-10mm}
        \caption{Current-controlled GFM.}
        \label{fig:current_control}
    \end{subfigure}
    \begin{subfigure}[b]{0.7\columnwidth}
    % % \vspace{-9mm}
        \centering
        \includegraphics[width=0.85\textwidth]{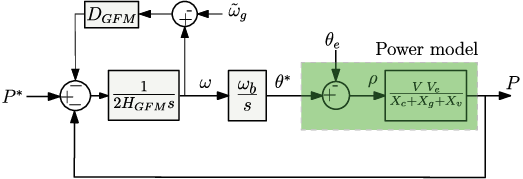}
        % \vspace{-10mm}
        \caption{Virtual synchronous machine (VSM-PLL).}
        \label{fig:VSM}
    \end{subfigure}
    \vspace{-1mm}
    \caption{GFM-VSC control details.}
    \label{fig:scheme_GFM}\vspace{-1mm}
\end{figure}

For the stability enhancement scenario, a GFM-VSC is connected at the PCC through a 30-km overhead transmission line (OHL), as illustrated in Fig.~\ref{fig:GFM_HVDC}. The GFM implements a current-control structure (Fig.~\ref{fig:current_control}) and a synchronization loop based on a Virtual Synchronous Machine (VSM) with a PLL (Fig.~\ref{fig:VSM}). Control and system parameters of the GFM-VSC are given in Tables \ref{tab:GFM_system} and \ref{tab:GFM_control}. For details on controller tuning and the linearized model, the reader is referred to \cite{paper_Anup}.

The analysis is organized in three parts: (A) validation of the LCC-HVDC impedance-based model connected to the grid; (B) impedance-based sensitivity analysis to determine the stability limits of the LCC-HVDC link under weak-grid conditions; and (C) impedance-based sensitivity analysis for stability enhancement of the LCC-HVDC system when a GFM-VSC is connected to a weak grid.

\begin{table}[t!]
\caption{GFM-VSC system parameters.}
\label{tab:GFM_system}
\centering
\begin{tabular}{ccc}
\toprule
\textbf{Component} & \textbf{Parameter} & \textbf{Value} \\
\midrule
\multirow{2}{*}{\parbox[c]{3cm}{\centering GFM}}
  & $S_{n,\mathrm{GFM}}$      & 1000\;MVA \\
  & $R_c / L_c$               & 0.005\;pu / 0.15\;pu \\
\addlinespace
\multirow{1}{*}{\parbox[c]{3.2cm}{\centering OHL}}
  & $R_{23} / L_{23}$         & 0.0072\;pu / 0.144\;pu \\
\bottomrule
\end{tabular}
\end{table}

\begin{table}[t!]
\caption{GFM-VSC control parameters.}
\label{tab:GFM_control}
\centering
\begin{tabular}{ccc}
\toprule
\textbf{Control loop} & \textbf{Parameter} & \textbf{Value} \\
\midrule
\multirow{2}{*}{\parbox[c]{3cm}{\centering VSM control}}
  & $D_{GFM} / H_{GFM}$ & 118.15 / 5\;s \\
  &                     & ($\zeta=1$) \\
\addlinespace
Virtual Impedance
  & $X_v$               & 0.3\;pu \\
\addlinespace
Filter
  & $\tau_{f_{CC}}$     & 40\;ms \\
\addlinespace
Current controller
  & $\omega_{CC}$       & 1200\;rad/s \\
\bottomrule
\end{tabular}
\vspace*{-4mm}
\end{table}

\subsection{Impedance-Based Model Validation}

For the validation of the impedance-based model of the system, the linear model introduced in \cite{paper_Anup} is employed, considering an operating point of $P = 1$ pu. For the FD identification and stability assessment, the open-source SIaD-Tool \cite{paper_SIaD} is used, placed between the LCC-HVDC and the grid, and applying a series-voltage perturbation scheme with a magnitude of 5\% of the nominal voltage.

A total of 114 logarithmically spaced frequency points are evaluated within the 1--1000 Hz range, using a 1 s time window, resulting in a frequency resolution of 1 Hz. The initialization time required to reach steady state is set to 15 s, and the settling time after each perturbation is fixed at 1.2 s. The discretization step is $\Delta t = 50\,\mu$s. The total execution time for FD identification and stability analysis at a single operating point averages 1506 s using MATLAB R2024b on a 1.6-GHz Intel i5 processor with 16 GB RAM. Further details on SIaD-Tool can be found in \cite{paper_SIaD}.

\begin{figure}[t!]
\centering
\includegraphics[width=0.85\columnwidth]{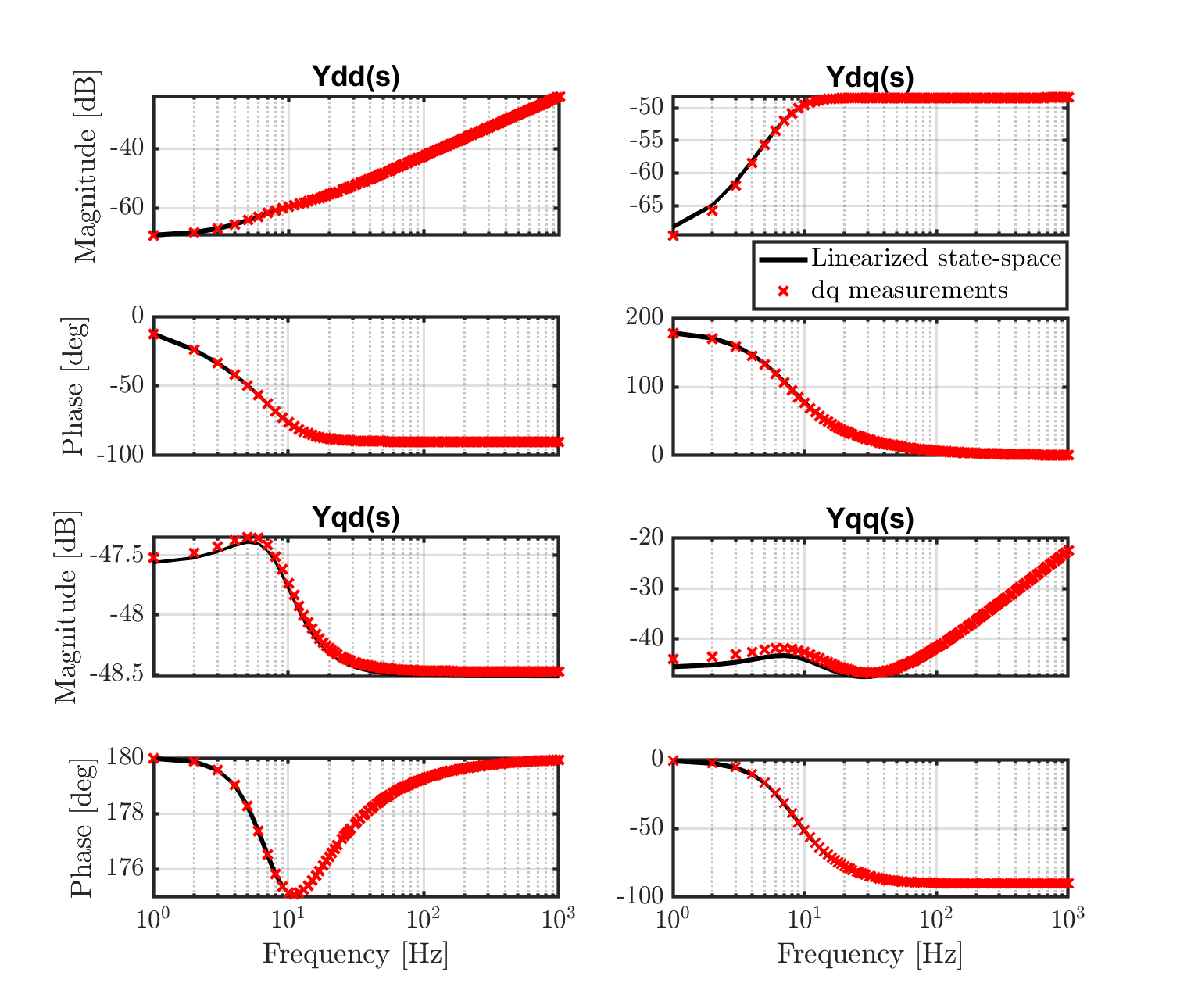}
\vspace{-4mm}
\caption{$dq$ frequency response of the LCC-HVDC link.}
\label{fig:Ydq_strong}
\vspace{-4mm}
\end{figure}

\begin{figure}[t!]
\centering
\includegraphics[width=0.75\columnwidth]{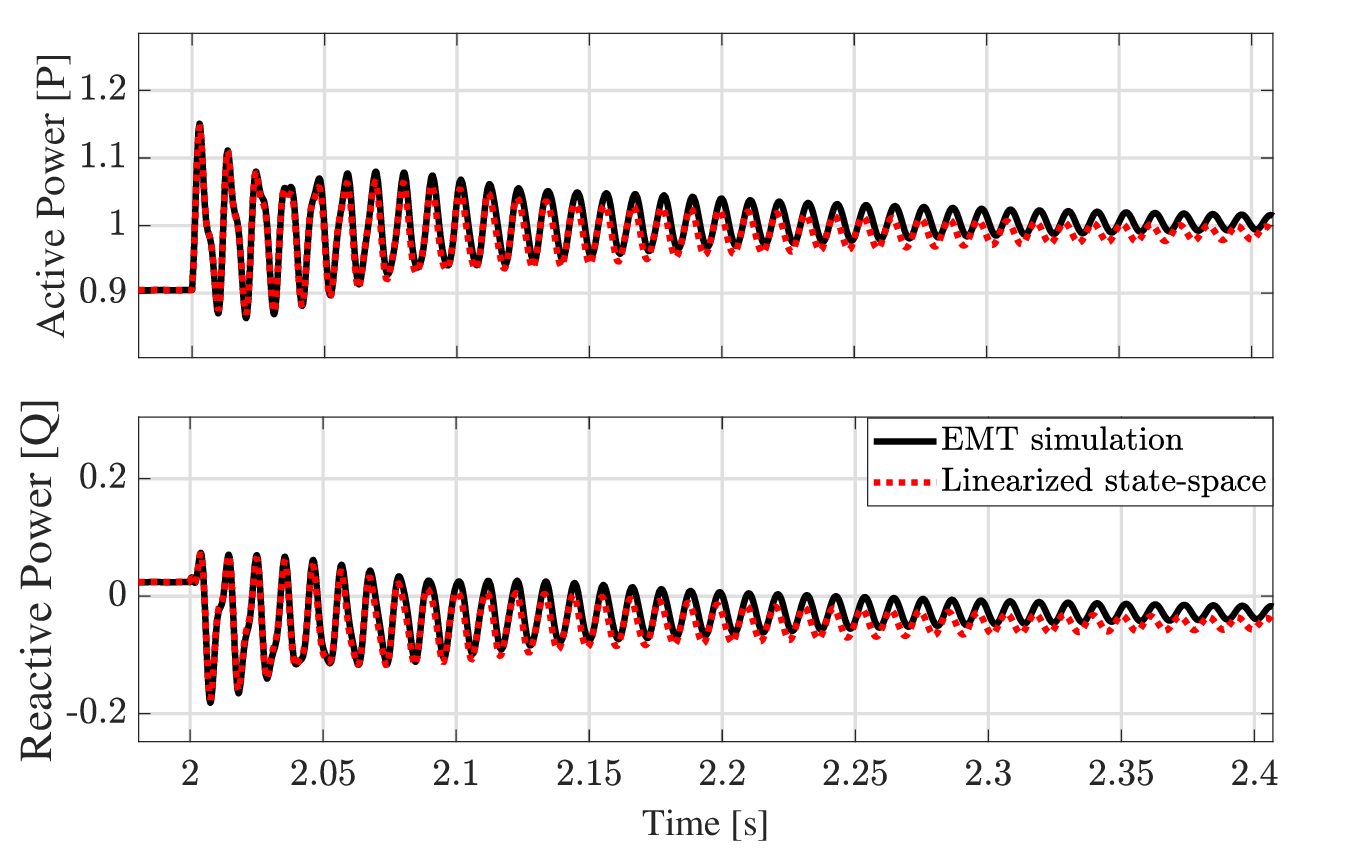}
\vspace{-2mm}
\caption{Time-domain model validation.}
\label{fig:time_domain}
\vspace{-2mm}
\end{figure}

Fig.~\ref{fig:Ydq_strong} presents the comparison between the open-loop linearized state-space response of the LCC-HVDC link and the identified $dq$ admittance model obtained with SIaD-Tool, considering an SCR of 10 (equivalent to $Z_g = 0.1$ pu). Additionally, Fig.~\ref{fig:time_domain} shows the time-domain validation comparing the linearized and the nonlinear EMT models after a 0.1 pu DC current step. Both figures exhibit excellent agreement, confirming the accuracy of the SSA-IB model.

Table~\ref{tab:modes_linear_strong} lists the modes of the closed-loop LCC-HVDC system connected to the grid. As no right-half-plane (RHP) poles are present, the system is confirmed to be stable, although two oscillatory modes with very low damping ($<$5\%) are observed.

\begin{table}[t!]
\centering
\caption{Theoretical modes under strong grid condition.}
\vspace{-1mm}
\label{tab:modes_linear_strong}
\resizebox{\columnwidth}{!}{
\begin{tabular}{cccccc}
\toprule
\textbf{Mode} & \textbf{ID} &  \textbf{Value [rad/s]} & \textbf{$\mathbf{\zeta}$ (\%)} & \textbf{f [Hz]} \\
\midrule
1, 2  & 3, 4 & - 8.0832 $\pm$ j960.99  & 0.841 & 153 \\ 
3 & 5 & - 12.937 + j0.0000  & 100.0 & 0.00 \\
4, 5  & 1, 2 & - 20.642 $\pm$ j1592.3  & 1.296 & 253 \\
6, 7  & 6, 7  & - 38.005 $\pm$ j35.623  & 72.96 & 5.67 \\
\bottomrule
\end{tabular}
}
\vspace{-5mm}
\end{table}

\begin{figure}[t!]
\centering
\includegraphics[width=0.7\columnwidth]{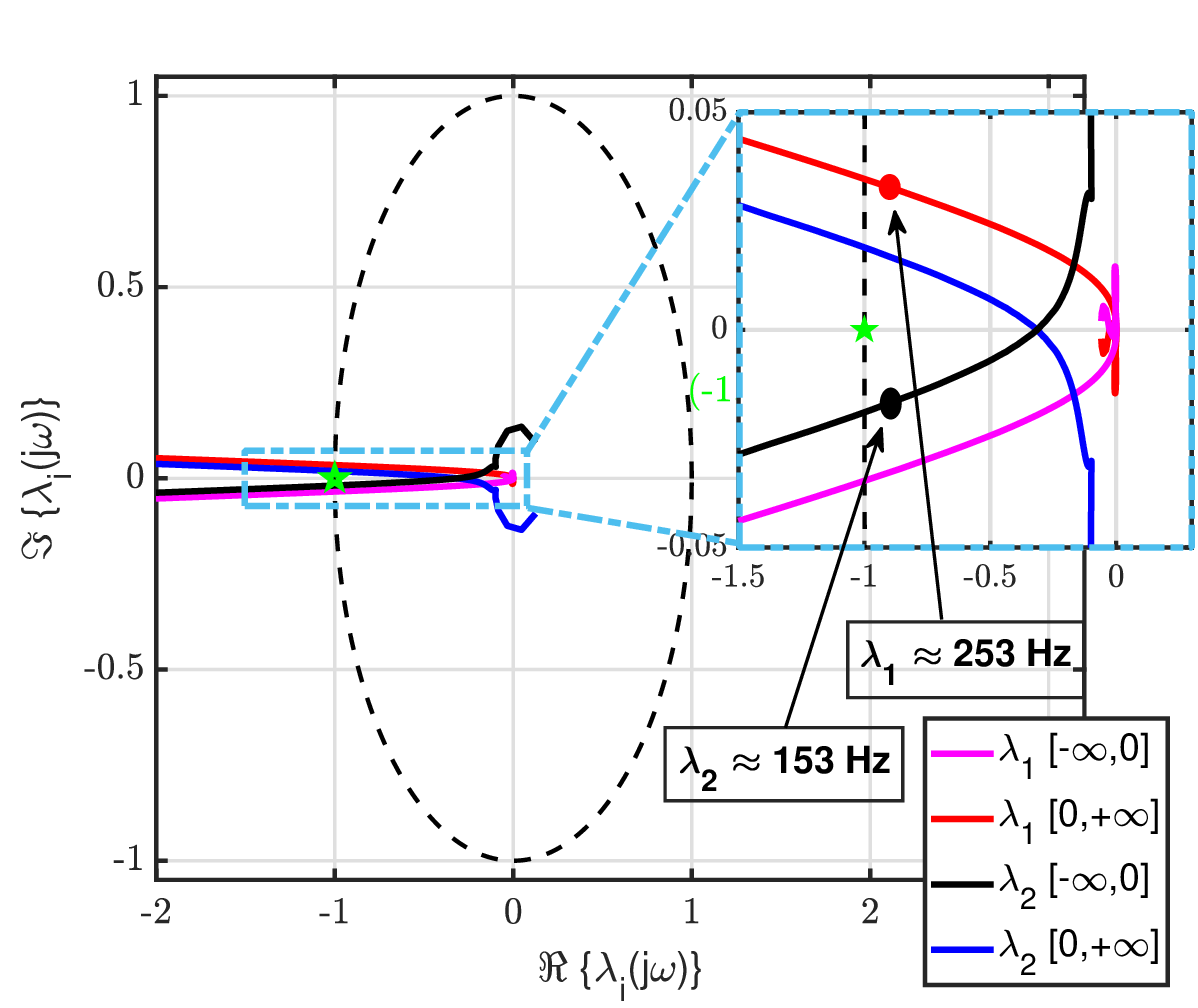}
\caption{GNC results for the closed-loop response.}
\label{fig:GNC_strong}
\vspace{-4mm}
\end{figure}

The closed-loop stability analysis using the impedance-based approach is performed through the GNC, with results shown in Fig.~\ref{fig:GNC_strong}. Since none of the eigenvalue loci of $\mathbf{L}(s)$ encircle the critical point $(-1,j0)$, the system is confirmed stable, consistent with the state-space modal analysis. The minimum distances of $\lambda_1$ and $\lambda_2$ to the critical point, located at 153 Hz and 253 Hz, respectively, reflect the critical oscillatory frequencies of the system.

\begin{figure}[t!]
% \vspace{-10mm}
    \centering
    \begin{subfigure}[b]{0.76\columnwidth}
        \centering
        \includegraphics[width=0.85\textwidth]{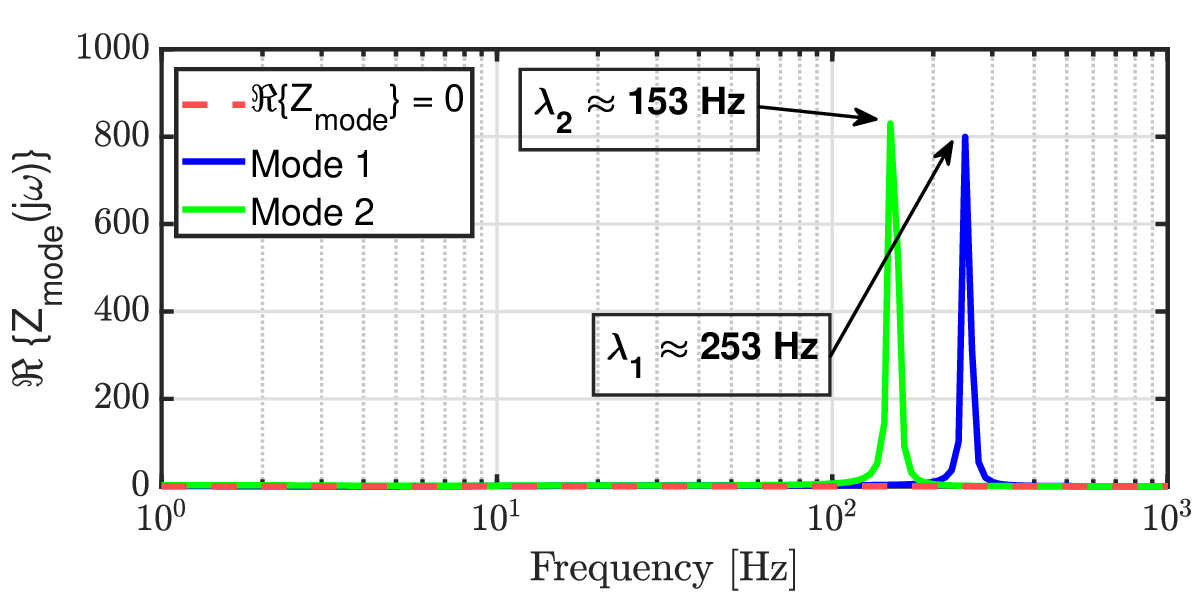}
        % \vspace{-10mm}
        \caption{Real part of modal impedances $Z_{mode}(s)$.}
        \label{fig:modal_strong_modes}
    \end{subfigure}
    \begin{subfigure}[b]{0.76\columnwidth}
    % % \vspace{-9mm}
        \centering
        \includegraphics[width=0.85\textwidth]{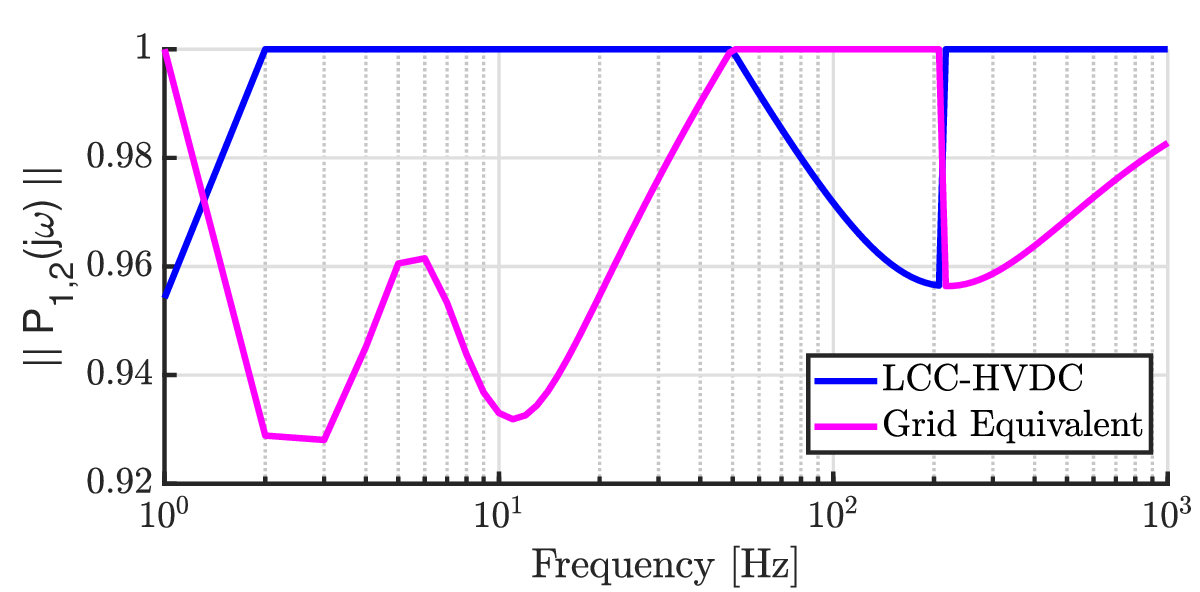}
        % \vspace{-10mm}
        \caption{Normalized participation factors $||P(s)||$.}
        \label{fig:modal_strong_PFs}
    \end{subfigure}
    % \vspace{-1mm}
    \caption{Impedance-based results of the system under study.}
    \label{fig:modal_strong}\vspace{-5mm}
\end{figure}

This is further verified through the modal impedances extracted from the closed-loop response $\mathbf{L}(s)$. Fig.~\ref{fig:modal_strong_modes} shows the peak values confirming the oscillatory modes of the complete system. As shown in Fig.~\ref{fig:modal_strong}, no negative modal impedances appear within the analyzed frequency range.

Finally, Fig.~\ref{fig:modal_strong_PFs} presents the participation factors of the eigenvalues of $L(s)$, confirming the origin of the oscillatory modes. The first mode is predominantly associated with the grid, while the second is linked to the LCC-HVDC link.

\subsection{Stability limit of the LCC-HVDC system}

\begin{figure}[t!]
\centering
\includegraphics[width=0.7\columnwidth]{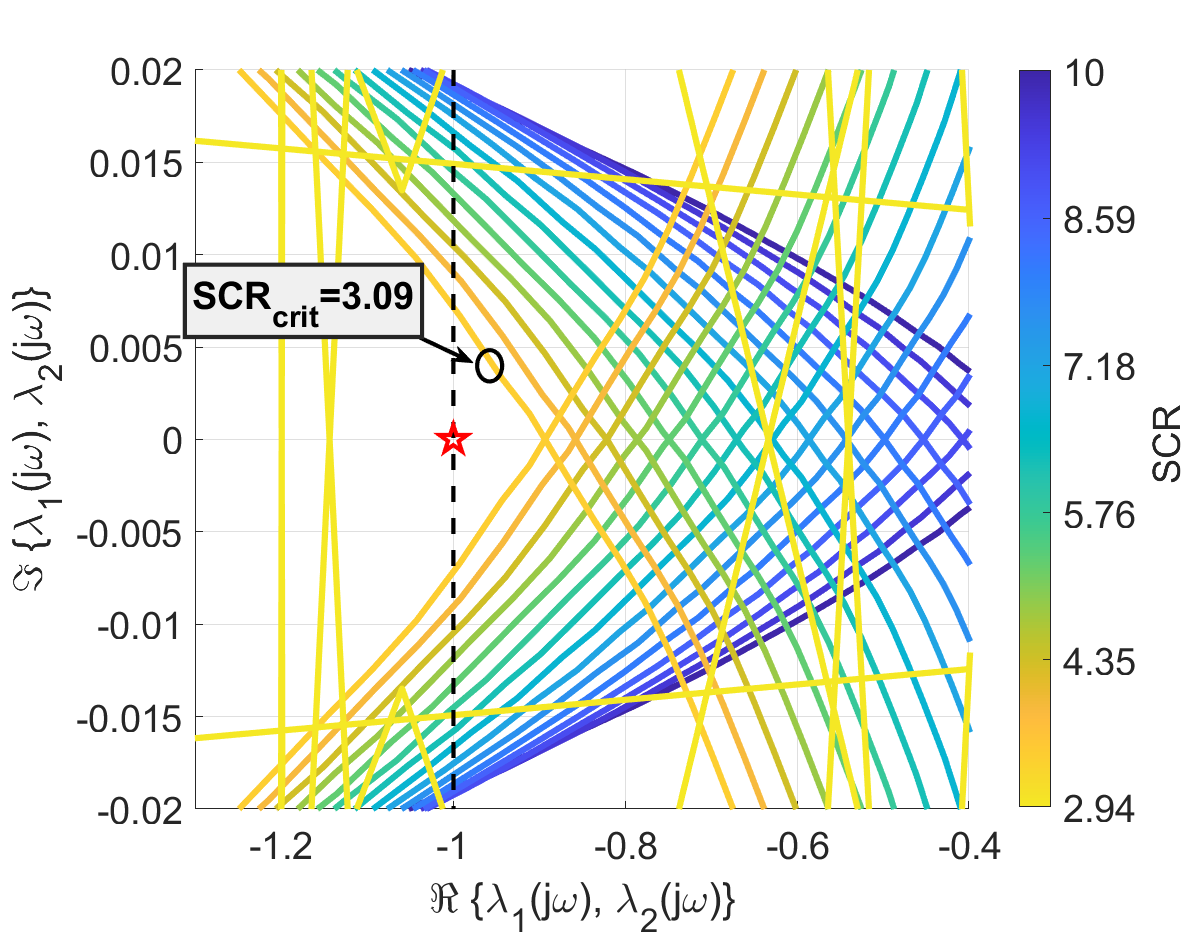}
\vspace{-2mm}
\caption{Sensitivity GNC results for the LCC-HVDC system.}
\label{fig:sensitivity_without_GFM}
\vspace{-4mm}
\end{figure}

The sensitivity analysis is first performed for the LCC-HVDC system without the GFM-VSC, in order to determine the stability limits of the system under weak grid conditions. This is performed by varying the impedance of Thévenin equivalent of the grid from an SCR of 10 to 2.94. In Fig. \ref{fig:sensitivity_without_GFM}, the results of the stability limit analysis are shown, where the limit is found with an SCR of 3.09. Notice that one value lower than 3.09 (the clearest yellow line) makes that the system eigenvalues loci change drastically their trajectories making several encirclements that can be seen graphically approaching from the left side.

\begin{figure}[t!]
% \vspace{-10mm}
    \centering
    \begin{subfigure}[b]{0.76\columnwidth}
        \centering
        \includegraphics[width=0.85\textwidth]{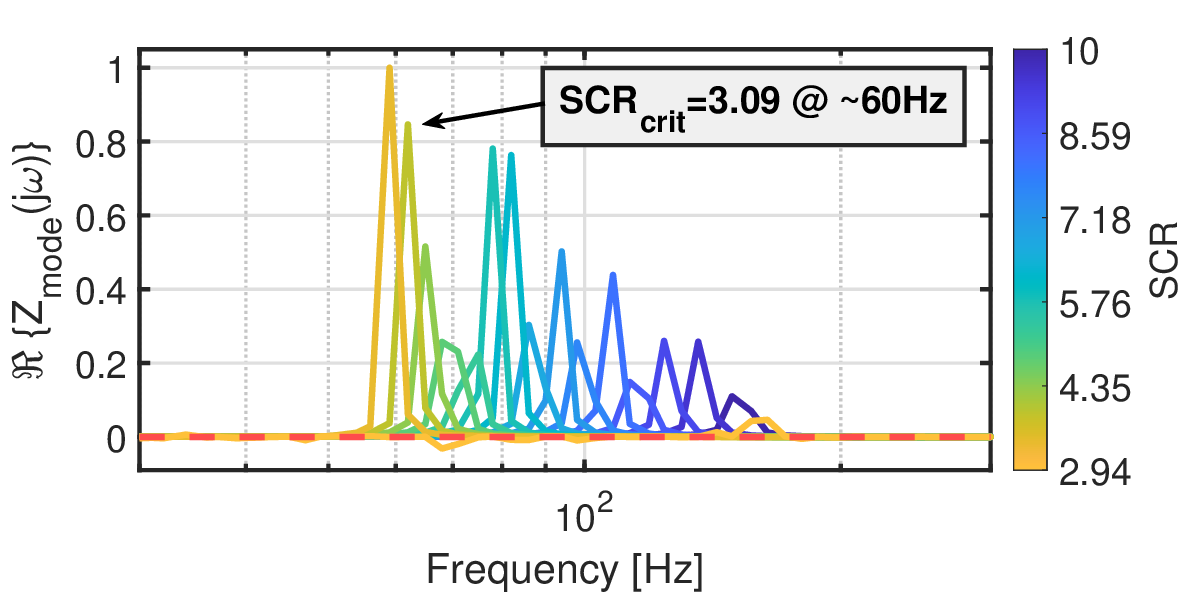}
        \vspace{-2mm}
        \caption{Mode 1 of the system.}
        \label{fig:mode1_LCC}
    \end{subfigure}
    \begin{subfigure}[b]{0.76\columnwidth}
    % % \vspace{-9mm}
        \centering
        \includegraphics[width=0.85\textwidth]{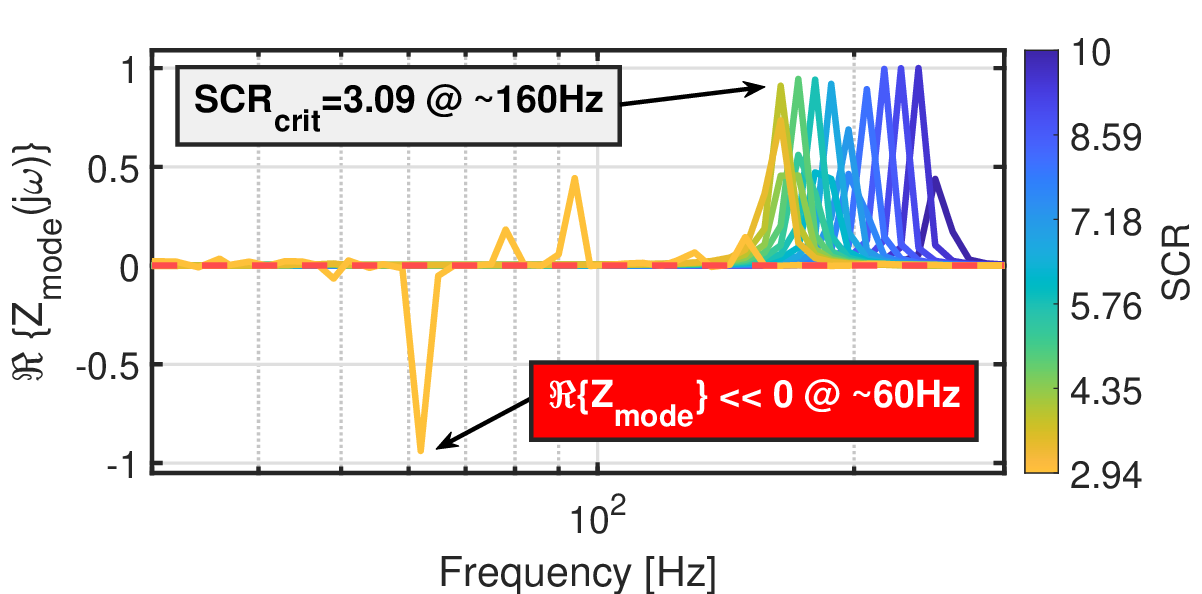}
        \vspace{-1mm}
        \caption{Mode 2 of the system.}
        \label{fig:mode2_LCC}
    \end{subfigure}
    \vspace{-2mm}
    \caption{Real part of modal impedances $Z_{mode}(s)$.}
    \label{fig:modes_sensitivity_LCC}\vspace{-1mm}
\end{figure}

\begin{table}[t!]
\centering
\caption{Critical modes for the LCC-HVDC system.}
\label{tab:modes_base_case}
\vspace{-1mm}
{
\begin{tabular}{cccccc}
\toprule
\multirow{2}{*}{\textbf{SCR}} 
& \multicolumn{2}{c}{ $\mathbf{Z_{mode1}}$ } 
& \multicolumn{2}{c}{ $\mathbf{Z_{mode2}}$ } \\
& \textbf{$\sigma_{norm}$} & \textbf{f [Hz]} 
& \textbf{$\sigma_{norm}$} & \textbf{f [Hz]} \\
\midrule
10.0  & 0.11 & 153 & 0.43 & 253 \\
8.62  & 0.26 & 139 & 1.00 & 241 \\
7.57  & 0.26 & 127 & 0.99 & 230 \\
6.75  & 0.15 & 116 & 0.99 & 220 \\
6.09  & 0.43 & 111 & 0.89 & 210 \\
5.55  & 0.25 & 101 & 0.46 & 200 \\
4.71  & 0.30 & 89  & 0.92 & 191 \\
4.38  & 0.76 & 85  & 0.94 & 183 \\
3.09  & 0.85 & 60  & 0.73 & 160 \\
2.94  & 1.00 & 55  & -0.93 & 60 \\
\bottomrule
\end{tabular}
}
\vspace{-6mm}
\end{table}

Fig. \ref{fig:modes_sensitivity_LCC} shows a zoom-in of the sensitivity displacement of the critical frequencies for the two modes of the whole system while varying the SCR. Here, the critical frequencies in the stability limits with an SCR of 3.09 of the LCC-HVDC connected to the grid are 60 Hz for the mode 1 and 160 Hz for the mode 2. These super-synchronous interactions are dominated by the grid and by the LCC, respectively, according to the participation factors analysis. When the SCR is reduced to 2.94, the mode 1 in Fig. \ref{fig:mode1_LCC} (bright yellow line) shows a minor negative $Z_{mode}(s)$ around 60 Hz and the mode 2 in Fig. \ref{fig:mode2_LCC} shows a bright negative value at 60 Hz, confirming a strong instability with a dominant oscillation frequency around 60 Hz, with high participation from the grid. Table~\ref{tab:modes_base_case} summarizes the critical modes for each SCR value. The normalized modal impedance used for comparison across all operating points (OPs) is defined as:
\begin{equation}\label{ec:norm}
\sigma_{\text{norm}} = 
\frac{\Re\{Z_{\text{mode}}(j\omega)\}}
     {\displaystyle \max_{\omega,\,\text{OP}} \big|\Re\{Z_{\text{mode}}(j\omega)\}\big|}
\end{equation}

\subsection{Stability Enhancement Using a GFM-VSC}

To improve the stability of the LCC-HVDC system under weak-grid conditions, a grid-forming VSC (GFM-VSC) is connected at the PCC according to the scheme in Fig.~\ref{fig:GFM_HVDC}. In the considered scenario, the GFM is providing only reactive power ($P_{GFM}=0$). The GFM-VSC rated power is set to 1000 MVA and a sensitivity analysis is performed.

\begin{figure}[t!]
\centering
\includegraphics[width=0.75\columnwidth]{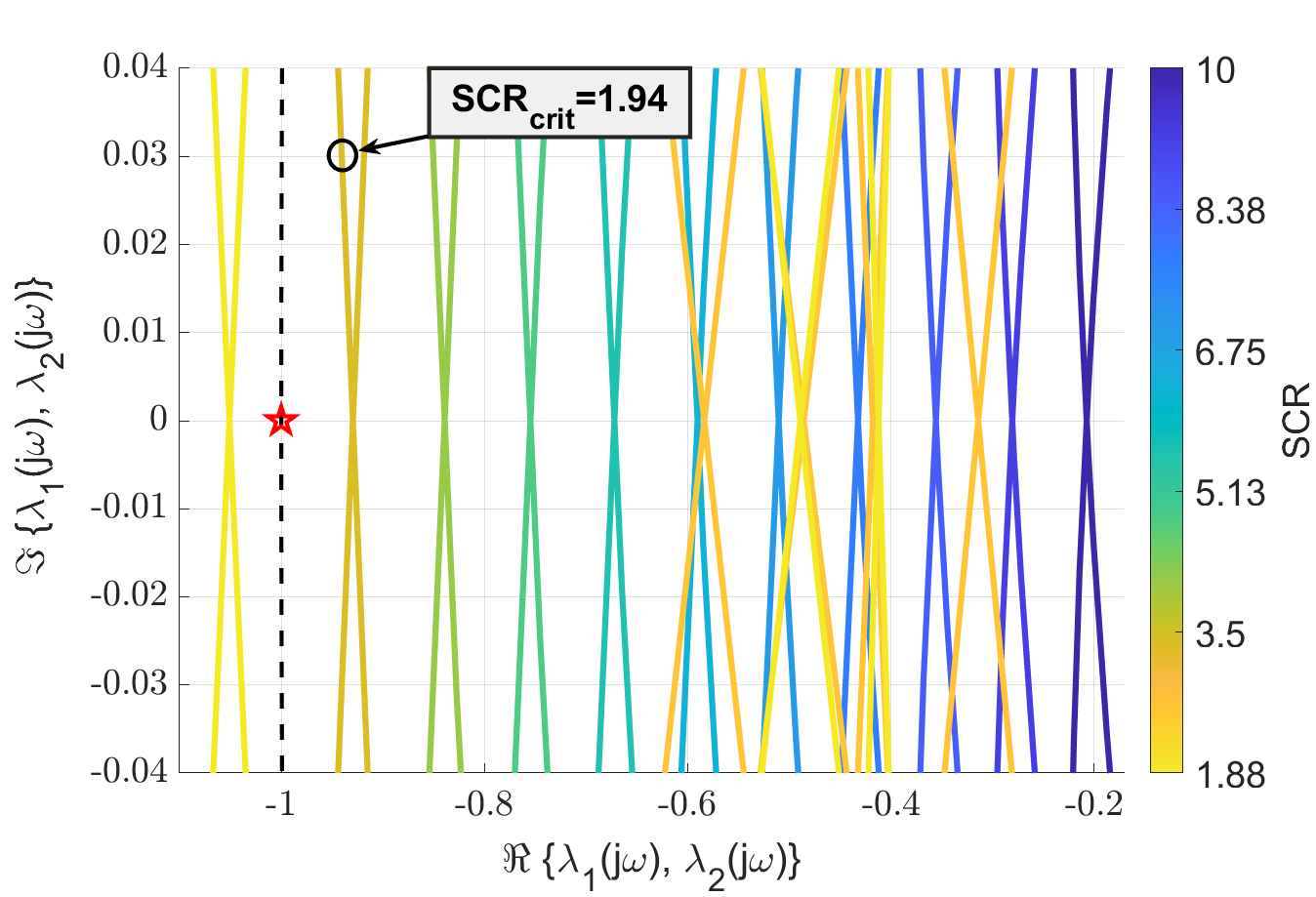}
\vspace{-2mm}
\caption{Sensitivity GNC results with the GFM-VSC.}
\label{fig:sensitivity_with_GFM}
\vspace{-4mm}
\end{figure}

\begin{figure}[t!]
% \vspace{-10mm}
    \centering
    \begin{subfigure}[b]{0.76\columnwidth}
        \centering
        \includegraphics[width=0.85\textwidth]{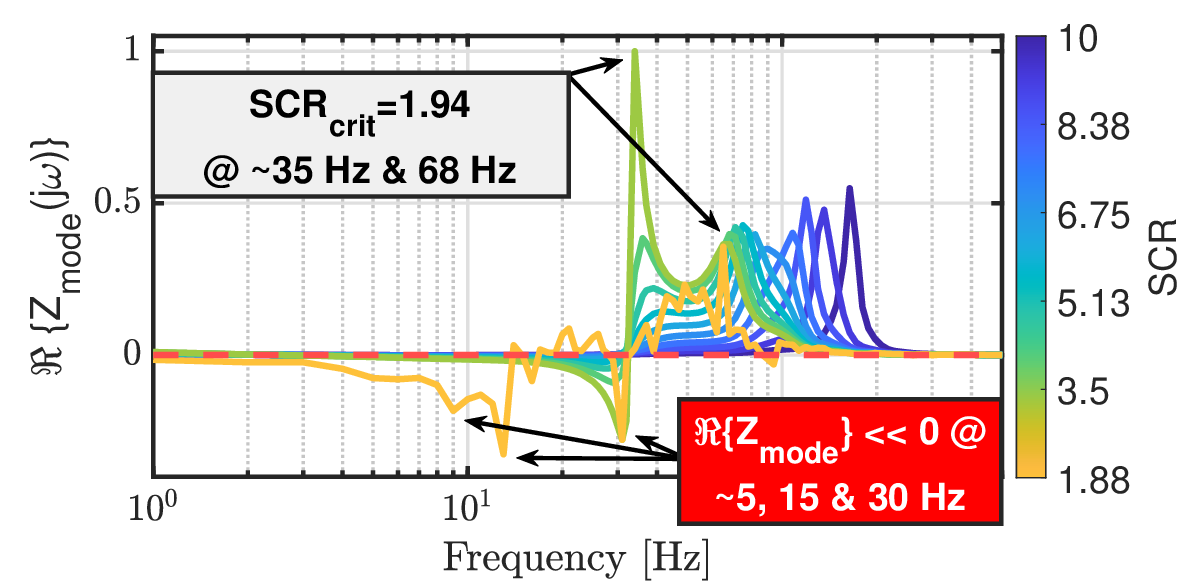}
        \vspace{-2mm}
        \caption{Mode 1 of the system.}
        \label{fig:mode1_GFM}
    \end{subfigure}
    \begin{subfigure}[b]{0.76\columnwidth}
    % % \vspace{-9mm}
        \centering
        \includegraphics[width=0.85\textwidth]{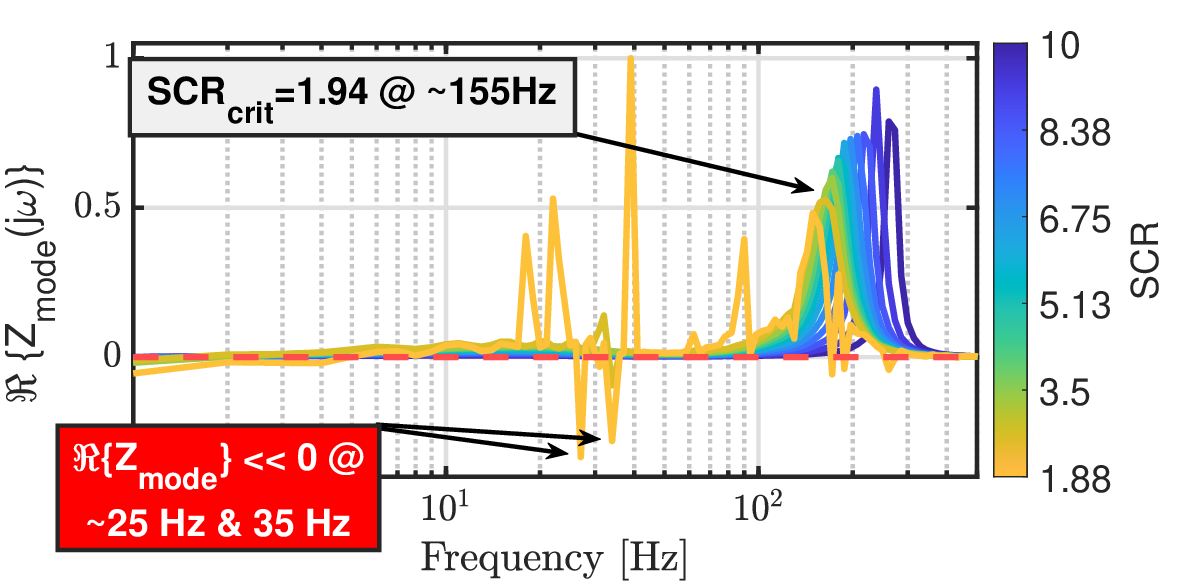}
        \vspace{-2mm}
        \caption{Mode 2 of the system.}
        \label{fig:mode2_GFM}
    \end{subfigure}
    \vspace{-2mm}
    \caption{Real part of modal impedances $Z_{mode}(s)$.}
    \label{fig:modes_sensitivity_GFM}\vspace{-2mm}
\end{figure}

\begin{table}[t!]
\centering
\caption{Critical modes with the GFM-VSC.}
\label{tab:modes_gfm_case}
\vspace{-1mm}
{
\begin{tabular}{cccccc}
\toprule
\multirow{2}{*}{\textbf{SCR}} 
& \multicolumn{2}{c}{ $\mathbf{Z_{mode1}}$ } 
& \multicolumn{2}{c}{ $\mathbf{Z_{mode2}}$ } \\
& \textbf{$\sigma_{norm}$} & \textbf{f [Hz]} 
& \textbf{$\sigma_{norm}$} & \textbf{f [Hz]} \\
\midrule
10.0  & 0.55 & 167 & 0.78 & 264 \\
7.35  & 0.47 & 139 & 0.89 & 241 \\
5.81  & 0.51 & 122 & 0.74 & 220 \\
4.80  & 0.40 & 111 & 0.73 & 210 \\
3.57  & 0.35 & 93  & 0.72 & 200 \\
3.16  & 0.39 & 85  & 0.71 & 191 \\
2.84  & 0.14, 0.42 & 44, 78  & 0.66 & 183 \\
2.17  & 0.38, 0.39 & 39, 71  & 0.62 & 175 \\
1.94  & 1, 0.36 & 35, 68  & 0.55 & 155 \\
1.88  & -0.32, -0.27 & 15, 30  & -0.33, -0.28 & 25, 35 \\
\bottomrule
\end{tabular}
}
\vspace{-4mm}
\end{table}

Fig.~\ref{fig:sensitivity_with_GFM} shows the SSA-IB stability-limit results with the GFM-VSC in place. The system remains stable down to an SCR of 1.94, representing a substantial improvement compared to the case without the GFM-VSC. For SCR values near 1.88, encirclements of the critical point begin to appear on the left side of the Nyquist plot, and additional trajectories become clearly visible on the right side, indicating instability.

Fig.~\ref{fig:modes_sensitivity_GFM} depicts the modal impedance evolution obtained from the sensitivity study. For mode 1 (Fig.~\ref{fig:mode1_GFM}), when SCR = 1.94 the critical frequencies appear at approximately 35~Hz and 68~Hz. A small negative excursion of the modal impedance is visible at SCR = 1.94; however, the GNC still indicates overall stability, which implies that the mode is very weakly damped around 30--35 Hz and therefore poses a potential sub-synchronous oscillatory-interaction risk dominated by the GFM-VSC. When the SCR is reduced to 1.88, mode 1 exhibits pronounced negative modal impedance peaks at about 5~Hz, 15~Hz and 30~Hz, indicating a strong instability dominated by low-frequency oscillations from the GFM-VSC side.

For mode 2 (Fig.~\ref{fig:mode2_GFM}), the critical frequency at SCR = 1.94 is near 155 Hz (grid-dominated). When the SCR decreases to 1.88, the dominant negative peaks shift to approximately 25 Hz and 35 Hz. Both modes present very negative modal impedances beyond the limit, confirming overall instability. Table~\ref{tab:modes_gfm_case} summarizes the critical modes for each SCR value.

\vspace{-1.1mm}
\section{Conclusions}

The frequency-domain, impedance-based sensitivity analysis methodology introduced in this work has proven to be an effective tool for assessing and enhancing the stability of LCC-HVDC systems operating under weak-grid conditions. The results demonstrate that the SSA impedance-based approach accurately reproduces the dynamic behavior of the system, even when detailed models are not available, thereby overcoming the limitations imposed by traditional SSA methods. The sensitivity analysis revealed that the LCC-HVDC link exhibits a stability limit at an SCR of approximately 3.09, dominated by a low-frequency grid-dominated oscillatory mode. The integration of a grid-forming VSC significantly increased the stability margin, shifting the limit to an SCR of 1.94 and favorably reshaping the modal distribution of the system. This behavior confirms the strong potential of GFM-VSCs as dynamic support devices in weak grids. Overall, the results validate the applicability of the SSA impedance-based framework for stability assessment in hybrid LCC-HVDC/VSC-GFM systems, offering a model agnostic and accurate methodology. Future work includes extending the analysis to more demanding dynamic scenarios, evaluating advanced GFM control strategies, and studying the coordinated operation of multiple support devices in highly weakened grid conditions.

\vspace{-1mm}
\section*{Acknowledgment}

This work has received funding from the ADOreD project under the European Union’s Horizon Europe Research and Innovation Programme under the Marie Skłodowska-Curie Grant Agreement No. 101073554.

\bibliographystyle{IEEEtran}
\bibliography{ref}

\end{document}